\documentclass[12pt,oneside, a4paper]{article}

\pdfoutput=1  
 
\ifx\pdfoutput\undefined
\usepackage[dvips,bookmarks=false]{hyperref}	
\else
\usepackage{hyperref}	
\fi
\hypersetup{colorlinks,bookmarksopen,bookmarksnumbered,citecolor=blue,
linkcolor=black,pdfstartview=FitH,urlcolor=blue}

\usepackage{graphicx}
\usepackage{subfigure}
\usepackage{amssymb}
\usepackage{cite}
\usepackage{bm}
\usepackage{amsmath,amsthm}
\usepackage{cleveref}

\numberwithin{equation}{section}

\newcommand{\gag}{g_{a\gamma}}

\begin{document}

\begin{titlepage}

\begin{center}

\vspace*{2cm}
        {\Large\bf Axion-electrodynamics:
          a quantum field calculation}
\vspace{1cm}

\renewcommand{\thefootnote}{\fnsymbol{footnote}}
{\bf Marc Beutter}\footnote[1]{marc.beutter@student.kit.edu}, 
{\bf Andreas Pargner}\footnote[2]{andreas.pargner@kit.edu}, 
{\bf Thomas Schwetz}\footnote[3]{schwetz@kit.edu}
				and  
{\bf Elisa Todarello}\footnote[4]{elisa.todarello@kit.edu} 
\vspace{5mm}

{\it%
{Institut f\"ur Kernphysik, Karlsruhe Institute of Technology (KIT),\\ 76021 Karlsruhe, Germany}
}

\vspace{8mm} 

\abstract{ An axion background field induces tiny oscillating electric
  and magnetic fields in an external static magnetic field. This
  signature is used to search for axion dark matter.  We use standard
  quantum field theory techniques to obtain an expression for a
  transition amplitude, from which we identify the classical
  electromagnetic fields induced by the background axion field.  We
  confirm previous results, that if the spatial size $R$ of the
  applied static magnetic field is small compared to the axion Compton
  wavelength $\lambda$, the induced electric and magnetic fields are
  parametrically suppressed by the small numbers $(R/\lambda)^2$ and
  $R/\lambda$, respectively, relative to the case when $R$ is larger than
  $\lambda$. Our approach allows an intuitive interpretation in terms
  of 4-momentum conservation and momentum exchange via the photon
  propagator.  }

\end{center}
\end{titlepage}

\renewcommand{\thefootnote}{\arabic{footnote}}
\setcounter{footnote}{0}

\setcounter{page}{2}

\section{Introduction}

The QCD axion \cite{Weinberg:1977ma, Wilczek:1977pj} is one of the
most attractive candidates for the dark matter (DM) in the Universe.  As
noted by Sikivie \cite{Sikivie:1983ip}, its coupling to the photon
field via the anomaly term offers experimental avenues to search for
galactic DM axions. A recent review about the
flourishing field of axion haloscopes can be found in
Ref.~\cite{Irastorza:2018dyq}.
The traditional way to search for axion DM is axion--photon conversion
in a resonant cavity, for example by the ADMX~\cite{Du:2018uak},
HAYSTAC~\cite{Zhong:2018rsr} or CULTASK~\cite{Woohyun:2016hkn} collaborations. 
These experiments explore the most
favored mass region for QCD axion DM, around
$10^{-6}-10^{-5}$~eV. Slightly larger masses can be reached by
dielectric haloscopes, such as MADMAX~\cite{Majorovits:2016yvk}.

More recently, some ideas to explore the region of axion masses
below $10^{-6}$~eV have been developed, see for instance
Refs.~\cite{Sikivie:2013laa,Kahn:2016aff} and the
ABRACADABRA~\cite{Ouellet:2018beu}, BEAST~\cite{McAllister:2018ndu}, 
DM Radio~\cite{Silva-Feaver:2016qhh} or KLASH~\cite{Gatti:2018ojx} projects. These searches are motivated in
the context of more general axion-like particle (ALP) DM
scenarios. While for the QCD axion both the axion mass $m_a$
and the axion--photon coupling $\gag$ are  related to the
same fundamental scale (the breaking scale of the Peccei--Quinn
symmetry \cite{Peccei:1977ur, Peccei:1977hh}), this needs not be
the case in general. For ALPs, $m_a$ and $\gag$ can be treated
as independent parameters. This is the approach we take here, and we use
the term ``axion'' in the more general sense, including also ALPs.

The axion--photon interaction is described by the Lagrangian
\begin{equation}\label{eq:Lag}
  \mathcal{L}_{a\gamma} = - \frac{\gag}{4} a F_{\mu\nu} \tilde
  F^{\mu\nu} = \gag a \vec{E}\cdot\vec{B} \,.
\end{equation}
Here, $a$ denotes the axion field, 
$F_{\mu\nu} = \partial_\mu A_\nu - \partial_\nu A_\mu$ is the
photon field-strength tensor and $\tilde F^{\mu\nu} = \frac{1}{2}
\epsilon^{\mu\nu\rho\sigma}F_{\rho\sigma}$ its dual. 
The coupling constant $\gag$ has dimension 1/energy.  In
the presence of this term, the inhomogenous Maxwell equations are
modified as \cite{Sikivie:1983ip,Sikivie:2013laa}
\begin{align}
  \vec{\nabla}\cdot\vec{E} &= \rho_e - \gag \vec{B}\cdot\vec{\nabla}a \label{eq:Maxwell1}\\
  \vec{\nabla} \times \vec{B} - \frac{\partial \vec{E}}{\partial t} &=
  \vec{j}_e - \gag\left(\vec{E} \times \vec{\nabla}a - \vec{B}\,\frac{\partial a}{\partial t}\right)\,,
   \label{eq:Maxwell2}
\end{align}
where $\rho_e$ and $\vec{j}_e$ are the external charge and current density, and
we have used Lorentz-Heaviside units with $c = 1$. 


Let us neglect axion-gradient terms and ignore the backreaction on the
axion field, which we treat as a classical source. Then the background
axion field in a homogeneous external magnetic field $\vec{B}^{\rm
  ext}$ induces an electric field,
\begin{align}\label{eq:Enaive}
  \vec{E}^{\rm ind} = -\gag a \vec{B}^{\rm ext} \,.
\end{align}
The basic approach of Refs.~\cite{Du:2018uak,Sikivie:2013laa,
  Kahn:2016aff, Ouellet:2018beu, McAllister:2018ndu,Zhong:2018rsr,
  Silva-Feaver:2016qhh,Woohyun:2016hkn,Majorovits:2016yvk} and similar
proposals is to search for those tiny induced electric or magnetic
fields, either by using broadband or resonant detection methods. Note
that ``tiny'' means that, for a DM axion with $m_a\sim
10^{-5}~\mathrm{eV}$ and a coupling $\gag\sim 10^{-15}~\mathrm{GeV}^{-1}$,
a $7~\mathrm{T}$ magnetic field induces an $E$ field with a magnitude
of the order $10^{-13}~\mathrm{V}/\mathrm{m}$.

Recently, there has been some controversy about the magnitude of the
induced electromagnetic (EM) fields in the case of an inhomogeneous
external magnetic field, more precisely, when the scale of
inhomogeneity is comparable to or smaller than the Compton wavelength of the
axion,
\begin{align}
  \lambda = \frac{2\pi \hbar}{m_a c} \approx 12\,{\rm m} \, \frac{10^{-7}\,{\rm eV}}{m_a c^2} \,.
\end{align}
Considering typical sizes of axion detection experiments, we see that
this question becomes relevant for $m_a c^2 \lesssim 10^{-6}$~eV.  As
discussed recently by Ouellet and Bogorad~\cite{Ouellet:2018nfr} and
Kim et al.~\cite{Kim:2018sci}, for experiments small compared to
$\lambda$ the induced electric field from \cref{eq:Enaive} is
suppressed by the small number $(R/\lambda)^2$, where $R$ is the
characteristic size of the external $B$-field used by the
experiment. This result emerges when the appropriate boundary
conditions are imposed on the solution of Maxwell's equations.  Tobar
et al.~\cite{Tobar:2018arx} come to a different conclusion, referring
to a quantum field theory calculation by Hong and
Kim~\cite{Hong:1991fp}.

In the following, we will add to this discussion by performing a
calculation of the axion--induced EM fields using quantum field
theory methods, similar to those used by Hong and Kim~\cite{Hong:1991fp}. By
properly taking into account the spatial extension of the external $B$
field via its Fourier transform, we obtain an intuitive understanding
of the three cases of ``large'', ``small'', and ``resonant''
experiments, characterized by $R \gg \lambda$, $R\ll \lambda$, and
$R\sim\lambda$, respectively, in terms of the available momentum
transfer. Our calculation confirms
the results of Refs.~\cite{Ouellet:2018nfr, Kim:2018sci}. Similar
methods have been used by Hill~\cite{Hill:2015kva,Hill:2015vma} to
calculate an induced oscillating electric dipole moment of the
electron and by Ioannisian et al.~\cite{Ioannisian:2017srr}
in the context of dielectric haloscopes.

Let us clarify the notion of Compton wavelength in this
context. In the situation of interest, namely detection of 
axion cold dark matter, we assume that the axion can be treated
as a classical background field which oscillates with the frequency $\omega
\approx m_a c^2/\hbar$. In a static external magnetic field, this
corresponds to a source for the induced EM field, which oscillates
with the frequency $\omega$. Therefore the induced EM radiation has a
wavelength $\lambda_{\rm EM} = 2\pi c/\omega \approx \lambda$. It is
the wavelength of this induced EM wave, which ultimately is
responsible for setting the relevant length scale to be compared with
$R$.  For a zero velocity axion field, it coincides with the quantum
mechanical Compton wavelength of an axion particle. In the following
we will continue to use the term axion Compton wavelength $\lambda$ to
denote this length scale, keeping in mind however, that the axion is a
classical field. From now on we will use natural units with $c=\hbar =
1$, such that $2\pi/\lambda = m_a \approx \omega$, where the
interpretation of $m_a$ as classical frequency is implied.

The outline of the paper is as follows. In \cref{sec:feynman}, we
calculate the induced fields using standard quantum field theory
methods to evaluate the relevant Feynman diagram. We discuss the
effect of gradient terms, and show that in the limit of
a homogeneous axion field, we recover the results of
Refs.~\cite{Ouellet:2018nfr, Kim:2018sci}, which were obtained by solving the classical
Maxwell's equations to leading order in $\gag$ and imposing the relevant
boundary conditions.  In \cref{sec:solenoid}, we provide an explicit
example by applying our method to the case of an infinitely
long solenoid. \Cref{sec:conclusion} contains a discussion of our
findings and the conclusions. In \cref{app:Hint} we derive the
relation between the interaction Lagrangian and Hamiltonian densities
for the axion--photon coupling, and in \cref{app:calc} we give technical details on the calculations of \cref{sec:solenoid}.

To fix the notation, greek indices take values 0,1,2,3, and roman ones
denote only the spatial components 1,2,3. For a four-vector we write
$(x^\mu) = (x^0,x^i) = (x^0, \vec{x})$, and our convention for the
Minkowski metric is $(g_{\mu\nu}) = \text{diag}(1,-1,-1,-1)$.

\section{Deriving the induced EM field}
\label{sec:feynman}

Let us first consider the interaction of a fermion,  say an
electron, with an external EM field:
\begin{equation}\label{eq:Lem}
\mathcal{L}_{\rm em}(x) = -e \overline{\psi}(x) \gamma^\mu\psi(x) A_\mu^{\rm ext}(x) \,.
\end{equation}
The transition amplitude for electron states
$|i\rangle \to |f\rangle$ in first order in perturbation theory is
\begin{equation}\label{eq:EMcurrent}
  \mathcal{A} = -i
  \langle f | \int d^4x \, \mathcal{H}_I(x) |i\rangle =
  -i\int d^4x J^\mu(x) A_\mu^{\rm ext}(x) \,,
\end{equation}
where the interaction Hamiltonian is given by $\mathcal{H}_I = -
\mathcal{L}_{\rm em}$ and we have defined $J^\mu \equiv e \langle f |
\overline{\psi} \gamma^\mu\psi |i\rangle$. Combined with the free
electron Hamiltonian, \cref{eq:EMcurrent} describes how an electron reacts to an external EM field.

\begin{figure}
  \centering
  \includegraphics[width=0.38\textwidth]{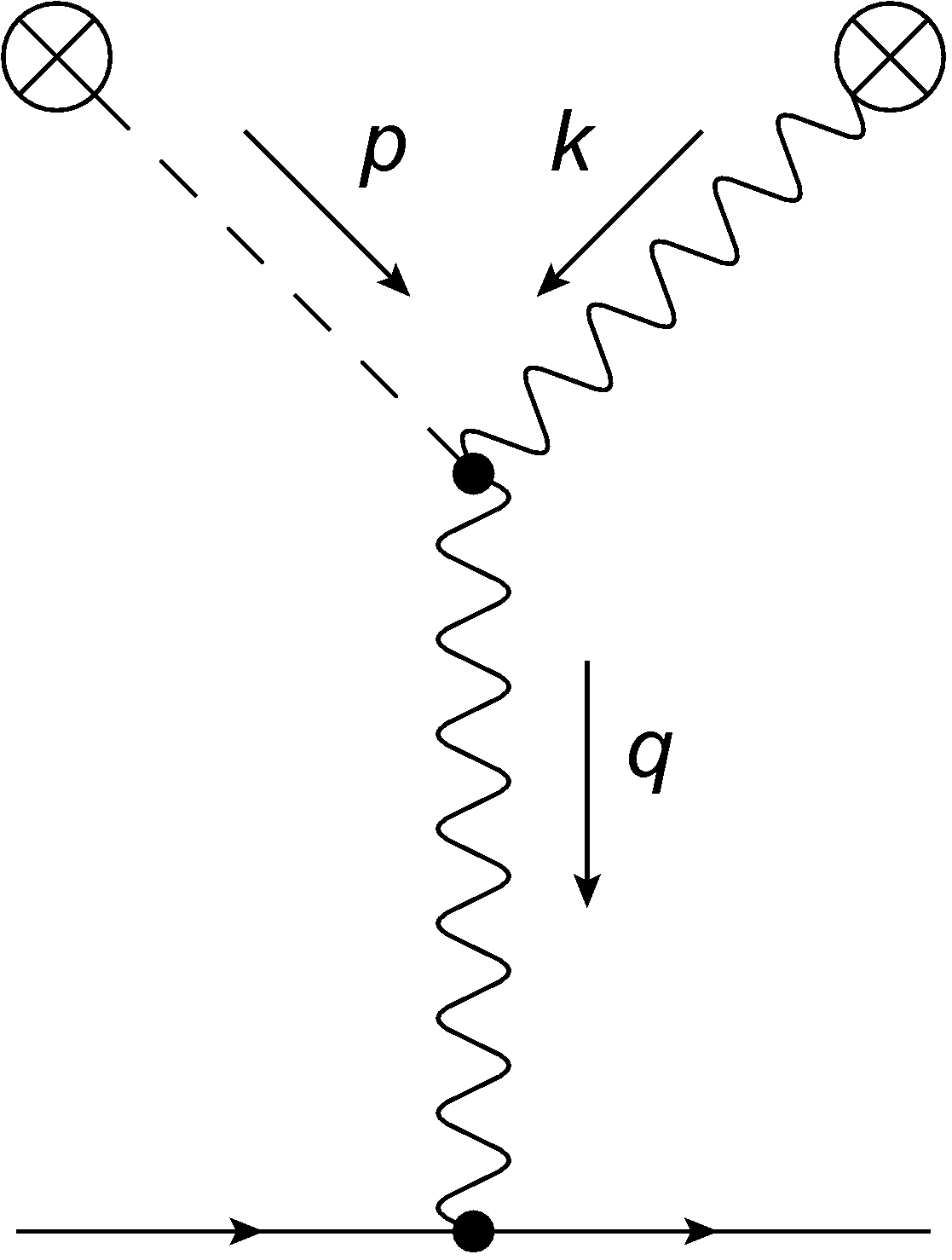}
  \caption{Interaction of an electron with an external EM field  and
    the axion DM field (dashed line) via virtual photon exchange. The $\otimes$ symbols indicate that the axion and the external EM field are treated as classical sources and are not affected by any backreaction.}
  \label{fig:feynman}
\end{figure}

Now we use a similar approach to calculate the effect of an external
EM field in combination with the axion dark matter background field on an electron,
based on the Feynman diagram in \cref{fig:feynman}.  The relevant
interaction Lagrangian is $\mathcal{L}_I = \mathcal{L}_{\rm em} +
\mathcal{L}_{a\gamma}$, with the axion--photon interaction given in
\cref{eq:Lag}. It can be shown that the interaction Hamiltonian density
can be written as $\mathcal{H}_I = - \mathcal{L}_I$ up to total
derivatives, despite the derivative coupling of the photon, see
\cref{app:Hint}.  The transition amplitude $|i\rangle \to |f\rangle$,
given the external axion and photon fields, can be calculated in second
order in perturbation theory:
\begin{equation}\label{eq:ampl}  
\mathcal{A} = \frac{(-i)^2}{2!}\langle f | T \int d^4x \int d^4y \, \mathcal{H}_I(x)  \mathcal{H}_I(y) |i\rangle \,,
\end{equation}
with $T$ denoting the time-ordered product. 

We use $a F_{\mu\nu} \tilde F^{\mu\nu} = 2 a (\partial_\mu A_\nu)
\tilde F^{\mu\nu} = - 2 (\partial_\mu a) A_\nu \tilde F^{\mu\nu}$ +
total derivative.  Then we contract $A_\nu$ with the photon field coupled
to the fermion current through the propagator. The other photon
field as well as the axion are treated as external sources. 
The amplitude corresponding to the diagram in \cref{fig:feynman} can be written as 
\begin{equation}\label{eq:JAind}
  \mathcal{A} = -i\int d^4x J^\mu(x) A_\mu^{\rm ind}(x) \,,
\end{equation}
where (including a combinatorial factor 2) 
\begin{equation}\label{eq:Aind_def}
  A_\mu^{\rm ind}(x) = i\gag \int d^4y D_{\mu\nu}(x-y)
  \partial_\rho a(y) \tilde F^{\rho\nu}(y) \,.
\end{equation}
The photon propagator in the Feynman gauge is given by
\begin{equation}\label{eq:D_F}
  D_{\mu\nu}(x-y) = \int \frac{d^4q}{(2\pi)^4} \frac{- i g_{\mu\nu}}{q^2 +i\epsilon}e^{-iq(x-y)} \,.
\end{equation}
We checked that physical quantities do not depend on the gauge
choice---as it should be, see also Ref.~\cite{Hill:2015vma}. By comparing
\cref{eq:JAind} with \cref{eq:EMcurrent}, we see that the combination
defined in \cref{eq:Aind_def} acts just as an external EM
field. Therefore we consider $A_\mu^{\rm ind}$ as given in
\cref{eq:Aind_def} as the induced vector potential seen by the
electron.

Let us now specify the properties of the external fields.  For the
external EM fields we assume that $\vec{E}$ as well as $\vec{B}$ are
constant in time. This implies for the Fourier transform of the field strength tensor
\begin{equation}\label{eq:Fext}
  \tilde F^{\rho\nu}(y) = \int \frac{d^3k}{(2\pi)^3} \tilde F^{\rho\nu}(\vec{k}) e^{i\vec{k}\cdot\vec{y}} \,.
\end{equation}
Furthermore, we require that the external EM fields fulfill the full set of zeroth-order Maxwell's
equations (without the axion term).

Galactic DM axions can be described as classical wave packets:
\begin{equation}\label{eq:ax}
  a(y) = a_0  \int \frac{d^3p}{(2\pi)^3} a(\vec{p}\,) e^{-ip y} 
\end{equation}
with $p^\mu = (\omega, \vec p)$. Since $a$ is a real field, it is understood that
we should take the real
part of \cref{eq:ax} as well as the following expressions.  Using that axions are
non-relativistic, we have $\omega \approx m_a + \vec{p}\,^2/2m_a$. The
group velocity of an axion wave packet is $v^i = d\omega/d p^i =
p^i/m_a$. The complex coefficients $a(\vec{p}\,)$ can be interpreted in
terms of a coarse-grained velocity distribution $f(\vec{v})$ by
$|a(\vec{p}\,)|^2 \propto f(\vec{v})$, see for example Ref.~\cite{Knirck:2018knd}. For galactic DM, the speed distribution is
peaked at $|\vec v| \sim 10^{-3}$ with a typical spread of similar
size, $\sigma_v \sim 10^{-3}$.  We adopt the normalization $\int
d^3p/(2\pi)^3 |a(\vec{p}\,)|^2 = 1$ and fix the constant $a_0$ by
assuming that the axion field provides the local dark matter density:
\begin{equation}\label{eq:rho_a}
  \rho_a \approx \frac{1}{2} m_a^2 a_0^2 \approx 0.4 \, {\rm GeV/cm^3} \,.
\end{equation}

Now we plug \cref{eq:D_F,eq:Fext,eq:ax} into \cref{eq:Aind_def}. The
$d^4y$ integration gives 4-momentum conservation via $(2\pi)^4\delta(q^0 -
p^0)\delta^{(3)}(\vec{k}+\vec{p} - \vec{q}\,)$. After shifting the
integration variable $\vec{q} \to \vec{q} + \vec{p}$ we obtain
\begin{align}\label{eq:Aind_gen}
  A_\mu^{\rm ind}(x) = -i\gag a_0 \int \frac{d^3q}{(2\pi)^3} \frac{d^3p}{(2\pi)^3} \,
  \tilde F_{\nu\mu}(\vec{q}\,) a(\vec{p}\,) \, \frac{p^\nu  \,
  e^{i\vec{q}\cdot\vec{x}} e^{-ip x}}{m_a^2 - \vec{q\,}^2 - 2\vec{q}\cdot\vec{p} + i\epsilon}  \,.
\end{align}

Note that this field is a classical object. In our derivation we
start from a relativistic covariant Lagrangian to construct a
quantum mechanical transition amplitude, \cref{eq:ampl}, using
standard quantum field theory ingredients, in particular the Feynman
propagator, \cref{eq:D_F}. However, in the comparison of our
expression for the amplitude with \cref{eq:EMcurrent} it is clear that
$A_\mu^{\rm ind}$ in \cref{eq:Aind_gen} is a classical
field. This is expected, since both the axion background field and the
external EM fields are classical sources.

\subsection{Discussion of the induced potential}

The expression \cref{eq:Aind_gen} relates the induced field to the
Fourier transform of the external EM field, showing how the induced
field depends on the geometry of the external field. Our approach
allows a physical interpretation in terms of the momentum transfer. In
particular, we identify the following three cases.
\begin{itemize}
\item {\it Large experiment:} If the size of the experiment, i.e.\ the
  region of non-zero external field, is large compared to $1/m_a$, the
  Fourier transform $\tilde F_{\mu\nu}(\vec{q}\,)$ is non-zero only for
  momenta $|\vec{q}\,| \ll m_a$. Hence, we can neglect all $\vec{q}$
  dependent terms in the denominator. This corresponds to the
  ``contact interaction'' with $q^2 \approx m_a^2$, as adopted in
  Ref.~\cite{Hong:1991fp}. In this limit we will recover the results
  obtained there.
\item {\it Small experiment:} The Fourier transform of the applied
  field extends to momenta much larger than $m_a$. The $q$-integral
  is dominated by the largest momenta and we can neglect $m_a^2 -
  2\vec{p}\cdot\vec{q}$ compared to $\vec{q}\,^2$ in the denominator. As we
  will show below, in this limit the potential will be suppressed by a
  factor $(m_a R)^2$.
\item {\it Resonant experiment:} If the size of the experiment is
  tuned to the Compton wavelength of the axion, the $q$ integral will
  be dominated by momenta $|\vec{q}\,|\sim m_a$. We clearly see the
  resonant enhancement in \cref{eq:Aind_gen}. In this case the precise
  shape of the applied EM field around momenta of order $m_a$ becomes
  important. Whether actually a resonance happens depends on
  additional boundary conditions. We will come back to this issue in
  \cref{sec:conclusion}.
\end{itemize}

We can write the induced potential, \cref{eq:Aind_gen} in terms of
the external $E$ and $B$ fields:
\begin{align}
  A_0^{\rm ind}(x) &= i\gag a_0 \int \frac{d^3q}{(2\pi)^3} \frac{d^3p}{(2\pi)^3} \,
  a(\vec{p}\,) \frac{e^{-i\omega t} e^{i(\vec{q}+\vec{p})\cdot\vec{x}}}
  {m_a^2 - \vec{q}\,^2 - 2\vec{q}\cdot\vec{p} + i\epsilon} \, \vec{p}\cdot\vec{B}^{\rm ext}(\vec{q}\,) \\
  \vec{A}^{\rm ind}(x) &= i\gag a_0 \int \frac{d^3q}{(2\pi)^3} \frac{d^3p}{(2\pi)^3} \,
  a(\vec{p}\,) \frac{e^{-i\omega t} e^{i(\vec{q}+\vec{p})\cdot\vec{x}}}
  {m_a^2 - \vec{q}\,^2 - 2\vec{q}\cdot\vec{p} + i\epsilon} \,
  \left[\omega\vec{B}^{\rm ext}(\vec{q}\,) + \vec{p}\,\times\vec{E}^{\rm ext}(\vec{q}\,)  \right]
  \label{eq:Aind}
\end{align}
with $t \equiv x^0$. In the last factors in these expressions, we
recognize the terms proportional to $\gag$ in the modified
Maxwell's equations, \cref{eq:Maxwell1,eq:Maxwell2}. We recover the well
known result, that an external static electric field couples to the axion
only via gradient terms. The behaviour in the limits $\lambda \ll R$
or $\lambda \gg R$ is the same for the terms proportional to $\omega$
and $\vec{p}$. Therefore the above mentioned suppression for small
experiments applies as well to momentum-dependent terms (here we
disagree with a corresponding remark in Ref.~\cite{Ouellet:2018nfr}).

Let us briefly discuss the impact of the axion velocity
distribution. Taking into account that the factor $a(\vec{p}\,)$
confines the $d^3p$ integral to the range $|\vec{p}\,| \sim 10^{-3}
m_a$, we observe that in the cases of large and small experiment the
$d^3p$ integral factorizes and the induced field becomes proportional
to
\begin{equation}
 a_0 \int \frac{d^3p}{(2\pi)^3}  p^\mu a(\vec{p}\,) e^{-ip x} = i \partial^\mu a(x) \,.
\end{equation}
Note that the induced potential is sensitive to the phases of
$a(\vec{p}\,)$, and therefore coherence effects may be relevant. A
discussion of those is beyond the scope of this work.

In contrast, for the resonant case, details of the axion velocity
distribution become relevant. Let us consider the $p^0$-term and
assume that $\tilde F_{\mu\nu}(\vec{q}\,)$ has such a shape that the
resonance dominates. Then we have approximately
\begin{align}
  \left. A_\mu^{\rm ind}(x) \right|_\text{res.} \approx i\gag a_0 \int_{|\vec{q}\,| \simeq m}
  \frac{d^3q}{(2\pi)^3}   \, \tilde F_{0\mu}(\vec{q}\,) e^{i\vec{q}\cdot\vec{x}}
  \int \frac{d^3p}{(2\pi)^3} \,
  \frac{a(\vec{p}\,)}{2 |\vec{p}\,| \cos\theta - i\epsilon} \,  e^{-ip x} \,,
\end{align}
where $\theta$ is the angle between $\vec{p}$ and $\vec{q}$. We see that
the finite momentum spread of the axions smears out the
resonance. This is a manifestation of the well known result that the
axion velocity spread leads to a quality factor of the resonance of
order $1/v^2 \sim 10^6$. We see that the result depends sensitively on
the shapes of $a(\vec{p}\,)$ and $\tilde F_{\mu\nu}(\vec{q}\,)$. Hence, in
principle an experiment whose size is comparable to the Compton
wavelength of the axion is sensitive to details of the local axion
velocity distribution, in addition to phase-coherence effects, see
Refs.~\cite{Foster:2017hbq,Knirck:2018knd}.

\subsection{Zero-velocity axions}

Let us from now on simplify the discussion by working in the approximation of axions
with zero velocity, $\vec{p} = 0$ and $\omega = m_a$. In this limit
the axion field becomes $a(x) = a_0 e^{-im_a t}$.
Then, the time component of the potential vanishes, $A_0^{\rm ind} =
0$. Using \cref{eq:Aind}, we find for the induced electric and
magnetic fields
\begin{align}
  \vec{E}^{\rm ind}(x) &= -\frac{\partial \vec{A}^{\,\rm ind}}{\partial t} =
  -\gag a_0 m_a^2 \, e^{-i m_a t}
  \int \frac{d^3 q}{(2\pi)^3} \, \frac{e^{i\vec{q}\cdot\vec{x}}}{m_a^2 - \vec{q}\,^2 + i\epsilon}
  \vec{B}^{\rm ext}(\vec{q}\,)  \,, \label{eq:Eind_0}\\
  \vec{B}^{\rm ind}(x) &= \vec\nabla \times \vec{A}^{\,\rm ind} =
  - \gag a_0 m_a \, e^{-i m_a t}
  \int \frac{d^3 q}{(2\pi)^3} \, \frac{e^{i\vec{q}\cdot\vec{x}}}{m_a^2 - \vec{q}\,^2 + i\epsilon}
  \vec{q}\times\vec{B}^{\rm ext}(\vec{q}\,)  \,. \label{eq:Bind_0} 
\end{align}

We can use \cref{eq:Eind_0,eq:Bind_0} to come back to the discussion
of the limiting cases in which the size of the experiment is either large
or small compared to the Compton wavelength of the axion. We assume
that the external magnetic field is contained mostly in a volume of
size $R$. This implies that the contribution to the momentum integral
will be dominated by momenta $|\vec{q}\,| \lesssim 1/R$. As mentioned
above, for the large experiment we can neglect $\vec{q}\,^2$ compared to
$m^2_a$ in the denominator. This corresponds to a point-like
interaction and we recover the ``naive'' result for the electric
field, \cref{eq:Enaive},
in agreement with Ref.~\cite{Hong:1991fp}. 

In the case of a small experiment, the integral has to be evaluated
including the pole. In order to get an estimate, we approximate the
magnetic field by a top-hat in momentum space, $\vec{B}^{\rm
  ext}(\vec{q}\,) \sim R^3 \vec{B}_0 \Theta(1/R - |\vec{q}\,|)$, with
$\vec{B}_0$ having the dimensions of a magnetic field. Let us further
consider the region in the center of the experiment $|\vec{x}| \ll R$,
which implies $|\vec{q}\cdot\vec{x}| \ll 1$. In this limit we
obtain\footnote{The integral can be evaluated by dividing it into a
  part up to momenta $\epsilon$ below the pole and from $\epsilon$
  above the pole and taking the limit $\epsilon \to 0$:
    \[
      \int_0^Q dq \frac{q^2}{m^2-q^2} = -Q + \frac{m}{2}\log\frac{Q+m}{Q-m} \approx -Q \,,
    \]
    where $Q \equiv 1/R > m$ and the last approximation holds for $Q\gg m$.}
  \begin{equation}
  \vec{E}^{\rm ind}(x) \simeq -\gag a_0 \, \vec{B}_0 e^{-i m_a t} (R m_a)^2 \,.
  \end{equation}
We see that the induced
electric field is suppressed by the small number $(R/\lambda)^2$.
Hence, our quantum field theory calculation confirms the
behaviour in the large and small wavelength limits obtained in
Refs.~\cite{Ouellet:2018nfr, Kim:2018sci} by solving the classical
Maxwell's equations. In \cref{sec:solenoid}, we demonstrate for a specific
field configuration that both approaches give quantitatively identical
results.

Before that, we can provide a further consistency check. Let us start
from \cref{eq:Eind_0} and plug in the inverse Fourier transform for
$\vec{B}^{\rm ext}$. Then the $d^3q$ integral can be performed,
yielding
\begin{equation}\label{eq:Eind1}
  \vec{E}^{\rm ind}(x) = 
  \gag a_0 m_a^2 \, e^{-i m_a t} \frac{1}{4\pi}
  \int d^3 y \, \frac{e^{im_a|\vec{x}-\vec{y}|}}{|\vec{x}-\vec{y}|}
  \vec{B}^{\rm ext}(\vec{y}\,)  \,.  
\end{equation}
Using further that the external $B$-field is generated by a current
$\vec{j}_e$ with $\vec{\nabla} \times \vec{B}^{\rm ext} = \vec{j}_e$
one can rewrite this expression as\footnote{One way to derive \cref{eq:Eind2} from
\cref{eq:Eind1} is to make use of the identity
\begin{equation*}
  \vec{\nabla}^2_y
  \frac{e^{im_a|\vec{x}-\vec{y}|}}{|\vec{x}-\vec{y}|} =
  -e^{im_a|\vec{x}-\vec{y}|}\left[4\pi\delta^3(\vec{x}-\vec{y})
   +\frac{m_a^2}{|\vec{x}-\vec{y}|} \right]
\end{equation*}
as well as
\begin{equation*}
\vec{B}^{\rm ext}(\vec{x}) = \frac{1}{4\pi} \int d^3y \,
\frac{\vec{\nabla} \times \vec{j}_e(\vec{y}\,)}{|\vec{x}-\vec{y}|} \,,
\end{equation*}
which follows from 
$\vec{\nabla}^2 \vec{B}^{\rm ext} = -\vec{\nabla} \times \vec{j}_e$.}
\begin{equation} \label{eq:Eind2}
  \vec{E}^{\rm ind}(x) = 
  \gag a_0 \, e^{-i m_a t} \frac{1}{4\pi}
  \int d^3 y \, \frac{e^{im_a|\vec{x}-\vec{y}|} -1}{|\vec{x}-\vec{y}|}
  \vec{\nabla} \times \vec{j}_e (\vec{y}\,)  \,.  
\end{equation}
Similarly, in \cref{eq:Bind_0} we can directly use 
$\vec{\nabla} \times \vec{B}^{\rm ext} = \vec{j}_e$ to obtain
\begin{equation} \label{eq:Bind2}
  \vec{B}^{\rm ind}(x) = 
  -i\gag a_0 m_a \, e^{-i m_a t} \frac{1}{4\pi}
  \int d^3 y \, \frac{e^{im_a|\vec{x}-\vec{y}|}}{|\vec{x}-\vec{y}|}
  \vec{j}_e (\vec{y}\,)  \,.  
\end{equation}
\Cref{eq:Eind2,eq:Bind2} agree with corresponding expressions in
Ref.~\cite{Ouellet:2018nfr} (up to a missing factor $1/4\pi$). There,
they have been derived as solutions of the Maxwell's equations with
the method of retarded Green's functions, confirming the equivalence
of the two methods. In this form we can also make direct contact with
standard EM radiation theory, see e.g., Chapter~9 of
Jackson~\cite{jackson_classical_1999}. We see that the limit of small
experiment is equivalent to the near-field approximation, amounting to
setting $e^{im_a|\vec{x}-\vec{y}|} \approx 1$ in integrals similar to
the ones in \cref{eq:Eind1,eq:Eind2,eq:Bind2}, see also
Ref.~\cite{Ouellet:2018nfr}.

\section{Infinitely long solenoid}
\label{sec:solenoid}

In this section, we consider the case of an external magnetic field generated by a
solenoid infinitely long in the $z$-direction and with radius
$R$. In this setup, we
use our method to calculate axion--induced EM
fields. In cylindrical coordinates, the external magnetic field 
reads
\begin{align}
\vec{B}^{\rm ext}(\vec{x},t) = B_0 \, \theta(R-r) \hat{e}_z \,.
\end{align}
Starting from \cref{eq:Aind} in the limit of zero momentum axions, we obtain
\begin{align}
A^z_\mathrm{ind}(r,z,t) &= i\gag a_0 B_0 e^{-im_a t} m_a R \int_0^\infty d q \, \frac{1}{m_a^2 - q^2 + i \epsilon} J_1(q R) J_0(q r) \,, 
\label{eq:Asolenoid}
\end{align}
where $J_n(x)$ is the Bessel function of the first kind of order $n$.
The integral can be solved analytically (see \cref{app:calc} for details):
\begin{align}
A^z_\mathrm{ind}(r,z,t) = i\gag a_0 B_0 e^{-im_a t}  m_a R^2
\begin{cases}
\displaystyle \frac{1}{m_a^2 R^2} - \frac{i \pi}{2} \frac{H_1^+(m_a R) J_0(m_a r)}{m_a R} & (r < R) \\
\displaystyle - \frac{i \pi}{2} \frac{J_1(m_a R) H_0^+(m_a r)}{m_a R} & (r > R)
\label{eq:A}
\end{cases} \,,
\end{align}
where $H^+_n(x)$ is the Hankel function of the first kind of order $n$.
The induced $E$ and $B$ fields resulting from \cref{eq:A} are 
\begin{align}
E^z_\mathrm{ind} &= - \gag a_0 B_0 e^{-im_a t} 
\begin{cases}
\displaystyle 1 - \frac{i \pi}{2} m_a R \, H_1^+(m_a R) J_0(m_a r) & (r < R) \\
\displaystyle - \frac{i \pi}{2} m_a R \, J_1(m_a R) H_0^+(m_a r) & (r > R)
\end{cases}
\,, \label{eq:Esolenoid}\\
B^{\phi}_\mathrm{ind} &= \frac{\pi}{2} \gag a_0 B_0 e^{-im_a t}  m_a R
\begin{cases}
\displaystyle H_1^+(m_a R) J_1(m_a r) &(r < R) \\
\displaystyle J_1(m_a R) H_1^+ (m_a r) & (r > R)
\end{cases} \,. \label{eq:Bsolenoid}
\end{align}
These results agree with the ones obtained in
Ref.~\cite{Ouellet:2018nfr} by solving the macroscopic equations of
motion for the induced EM field for the same configuration. In this
case, the boundary conditions that have to be imposed explicitly to
solve Maxwell's equations in Ref.~\cite{Ouellet:2018nfr} are already
contained in our approach. They are encoded in the Fourier transform
of the static magnetic field.

In the limit of an experiment small compared to the Compton wavelength
($R m_a \ll 1$) we get
\begin{align}
E^z_\mathrm{ind} &\approx \begin{cases}
\displaystyle \frac{1}{4} \gag a_0 B_0 e^{-im_a t}  (m_a R)^2 \left(1 + i \pi - 2 \gamma - 2 \log \left(\frac{m_a R}{2} \right) - \left(\frac{r}{R} \right)^2 \right) & (r < R \ll \lambda) \\
\displaystyle \frac{1}{4} \gag a_0 B_0 e^{-im_a t}  (m_a R)^2 \left(i \pi -2 \gamma - 2 \log \left(\frac{m_a r}{2} \right) \right) & (R < r \ll \lambda)
\end{cases}
\,, \label{eq:Esol_small}\\
B^{\phi}_\mathrm{ind} &\approx \begin{cases}
\displaystyle \frac{i}{2} \gag a_0 B_0 e^{-im_a t}  m_a r &(r < R \ll\lambda) \\
\displaystyle \frac{i}{2} \gag a_0 B_0 e^{-im_a t}  m_a \left(\frac{R^2}{r} \right) &
(R < r \ll\lambda)
\end{cases} \,,
\end{align}
with $\gamma$ being the Euler-Mascheroni constant. Again,
the induced magnetic field is suppressed by a factor
$R / \lambda$ and the electric field by a factor $(R / \lambda)^2$ for
$R \ll \lambda$. Note that for constant DM energy density,
\cref{eq:rho_a} implies that $a_0 \propto 1/m_a$. Therefore, in the
small experiment limit, the electric field inside the experiment
scales as $E^z_\mathrm{ind} \propto m_a R^2$, while
$B^\phi_\mathrm{ind} \propto r$ ($r<R$) and is independent of $m_a$.
In the limit $R \gg \lambda$, the spatial average of the electric field
approaches the ``naive" solution $E \approx -\gag a_0 B_0 e^{-im_a t} $ and the
averaged magnetic field vanishes.

Let us mention that our results agree with Ref.~\cite{Tobar:2018arx}
for the magnetic field in the region outside the solenoid, $r > R$, 
while we disagree for the electric field in the inside region, $r <
R$ in the case of small experiment.

\section{Discussion and conclusion}
\label{sec:conclusion}

To summarize, motivated by recent experimental developments,
Refs.~\cite{ Ouellet:2018beu, McAllister:2018ndu,
  Silva-Feaver:2016qhh}, we have calculated the EM fields induced by
dark matter axions in the presence of a static EM field by using
quantum field theory methods. We apply standard techniques to obtain
an expression for a transition amplitude describing the interaction of
an electron with the background axion and EM fields. The corresponding
Feynman diagram is shown in \cref{fig:feynman}. We identify the
effective vector potential to which the electron current couples.  The
internal photon line is described via the Feynman propagator. Indeed,
it is the shape of this propagator, together with 4-momentum
conservation, which determines the behaviour of the induced field in
the three regimes of ``small'', ``large'' or ``resonant''
configurations.  Using the Fourier transform of the external EM field,
we obtain an intuitive interpretation in terms of the available
momentum modes which the external field can provide, depending on its
spatial shape. If the field has a size $R$ large compared to the
Compton wavelength $\lambda$ of the axion, the available momenta are
small compared to the axion mass $m_a$, and can be neglected in the
propagator and we obtain the contact-interaction limit adopted, e.g.,
in Ref.~\cite{Hong:1991fp}. In the limit of a small experiment
($R\ll\lambda$), the propagator is dominated by the large momenta of
the external field, which naturally lead to a suppression of the
induced fields by two (one) powers of $R/\lambda$ for the electric
(magnetic) field.

Our results for the induced EM fields agree with the ones from
Refs.~\cite{Ouellet:2018nfr, Kim:2018sci}. However, we find that the
results for the induced electric field inside the experiment obtained
in Ref.~\cite{Tobar:2018arx}~(arXiv v4) do not apply in the case of
the ``small'' experiment, while Ref.~\cite{McAllister:2018ndu} assumes
they do. The calculations in Ref.~\cite{Ouellet:2018nfr,
  Kim:2018sci} are based on classical solutions of Maxwell's
equations. In this traditional approach the appropriate boundary
conditions of all involved fields are essential to obtain the correct
behaviour in the case of a ``small'' experiment. Our calculation shows
that consistent results can be obtain also by a quantum field theory
calculation in terms of the available 4-momentum flow for the virtual
photon mediating the interaction.

Let us briefly comment on the resonant case, when the spatial size of
the applied field becomes comparable to the axion Compton
wavelength. In this case, the detailed shape of the field
configuration becomes important. Note that an actual resonant
enhancement generally requires additional boundary conditions, beyond
the ones encoded in the shape of the applied magnetic field. This can
be seen from the explicit calculation for the infinitely long solenoid
in \cref{sec:solenoid}: indeed, \cref{eq:Esolenoid,eq:Bsolenoid} do
not show any resonance as a function of $R$ for fixed $m_a$. The
reason is that, in a configuration as in a resonance cavity or with
dielectric layers, additional boundary conditions have to be imposed
directly on the photon emitted at the axion-vertex. The appropriate
quantum field theory methods to describe such a process have been
presented in Ref.~\cite{Ioannisian:2017srr}. With our ansatz we assume
that the induced field is a virtual photon described by the standard
Feynman propagator, \cref{eq:D_F}, coupling to a fermion current which serves as
``detector''.

Finally let us stress, that although we start from a relativistic
quantum field theory expression for a transition amplitude, the
induced EM field we obtain is classical. While most of the results
obtained here are known in the axion literature, our approach offers
an alternative derivation and additional physics insights. Employing
standard quantum field theory methods, the calculations presented
above provide some clarification in the recent discussion about
axion--induced EM fields for experiments exploring the region of small
axion masses, below $10^{-7}$~eV.

\subsection*{Acknowledgments}

This project is supported by the European Union’s Horizon
2020 research and innovation programme under the Marie
Sklodowska-Curie grant agreement No 674896 (Elusives).
A.P. acknowledges the support by the DFG-funded Doctoral School KSETA.

\appendix

\section{The interaction Hamiltonian}
\label{app:Hint}

The interaction Hamiltonian can be obtained by the standard canonical
formalism, see e.g. Ref.~\cite{weinbergQFT}. Let us consider the following
Lagrangian density for the photon field
\begin{equation}\label{eq:L}
  \mathcal{L} =  -\frac{1}{4} F_{\mu\nu} F^{\mu\nu} 
  -\frac{\gag}{4} a F_{\mu\nu} \tilde F^{\mu\nu} 
  -J^\mu  A_\mu \,,
\end{equation}
which is a function of the photon field $A_\mu$ and its
derivatives. The conjugate momentum is obtained as
\begin{equation} \label{eq:pi}
  \pi_\mu \equiv \frac{\delta \mathcal{L}}{\delta \dot{A}^\mu} =
  - F_{0\mu} - \gag a\tilde F_{0\mu} \,,
\end{equation}
where the dot denotes derivative with respect to time.
We find that $\pi_0 \equiv 0$, which implies that $A^0$ is not a
canonical variable and can be eliminated by using equations of motion.
The Hamilton density is obtained by a Legendre transform of the Lagrangian density
\begin{equation}
  \mathcal{H} = \pi_\mu \dot{A}^\mu - \mathcal{L} \,,
\end{equation}
where time derivatives of the canonical variables are expressed by
their conjugate momenta. Hence we have to replace $F_{0i}$ by $\pi_i$
using \cref{eq:pi}. Note that $\tilde F_{0i}$ only contains the
spatial components $F^{jk}$ and therefore no time derivative. From \cref{eq:pi} we have
$\partial^0A^\mu = - \pi^\mu + \partial^\mu A^0 - \gag a \tilde F^{0\mu}$, which gives
\begin{align}
  \mathcal{H} &= -\pi_i\pi^i + \pi_i\partial^iA^0 - \gag a \pi_i\tilde F^{0i} \\
  &+ \frac{1}{2}F_{0i}F^{0i} + \frac{1}{4}F_{ij}F^{ij} + \gag a F_{0i}\tilde F^{0i} + J^\mu A_\mu \,.
\end{align}
In the last line we have used $F_{\mu\nu}\tilde F^{\mu\nu} = 4
F_{0i}\tilde F^{0i}$. Now we use \cref{eq:pi} to eliminate $F_{0i}$ in
the last line. Keeping terms up to linear order in $\gag$ we find
\begin{align}
  \mathcal{H} &= \frac{1}{2} \pi^i\pi^i + \frac{1}{4}F_{ij}F^{ij} - \gag a \pi_{i}\tilde F^{0i}
  \label{eq:H1}\\
  &+ \pi_i\partial^iA^0 + J^\mu A_\mu \,.
  \label{eq:H2}
\end{align}
The second line can be re-written by using the equation of motion
following from the Lagrangian \cref{eq:L}:
\begin{equation}
  \partial_\nu F^{\nu\mu} + \gag\partial_\nu(a \tilde F^{\nu\mu}) - J^\mu = 0 \,.
\end{equation}

Comparing the zero-component with \cref{eq:pi} we find
$\partial_i\pi^i =  J^0$.  Hence, the terms in \cref{eq:H2} can be
written as $J^iA_i$ + total derivative, and we have eliminated $A^0$
from the Hamiltonian.

In summary, the first two terms in \cref{eq:H1} can be identified with
the Hamiltonian of the free electro-magentic field, while the last
term is just the negative of the interaction Lagrangian,
\cref{eq:Lag}.  Hence, we arrive at the result that the interaction
Hamiltonian is given by $\mathcal{H}_I = -\mathcal{L}_I$ up to linear
order in $\gag$.

\section{Integrals needed for the infinitely long solenoid calculation}
\label{app:calc}

We start from \cref{eq:Asolenoid} and perform the substitution $q\to k/R$. To evaluate the induced vector potential, we need to solve the integral
\begin{align}
I(m_a,r,R) \equiv\int_0^\infty d k \, \frac{1}{m_a^2 R^2 - k^2 + i \epsilon} J_1(k) J_0\left(k \frac{r}{R}\right) \, .
\end{align}
This integral has a complicated structure but can be reduced to expressions known in the literature. For this purpose, we have to differentiate between the cases $ r > R$ and $r < R$:
\begin{enumerate}
\item $r > R$: using the Sokhotski-Plemelj theorem, we obtain
\begin{align}
I(m_a,r,R) &= \mathcal{P} \int_0^\infty d k \, \frac{1}{m_a^2 R^2 - k^2} J_1(k) J_0 \left(k \frac{r}{R} \right) + i \pi \mathrm{Res} \left(\frac{1}{m_a^2 R^2 - k^2} J_1(k) J_0 \left(k \frac{r}{R} \right) \right) \enspace,
\end{align}
where $\mathcal{P}$ denotes the Cauchy principal value.  With the help of Ref.~\cite[p.~463~(57)]{Bessel2}
\begin{align}
\mathcal{P} \int_0^\infty d k \, \frac{1}{m_a^2 R^2 - k^2} J_1(k) J_0 \left(k \frac{r}{R} \right) = \frac{\pi}{2} \frac{J_1(m_a R) Y_0(m_a r)}{m_a R}  \enspace,
\end{align}
we get
\begin{align}
I(m_a,r,R) = - \frac{i \pi}{2} \frac{J_1(m_a R) H_0^+(m_a r)}{m_a R} \, .\label{I1}
\end{align}

\item $r < R$: we use integration by parts
\begin{align}
I(m_a,r,R) = &\lim_{\epsilon \rightarrow 0^+} \left[ \frac{1}{m_a^2 R^2 - k^2 + i \epsilon} J_0(k) J_0 \left(k \frac{r}{R} \right) \right]_0^\infty \nonumber \\
+ &\lim_{\epsilon \rightarrow 0^+} \int_0^\infty d k \frac{2k}{(m_a^2 R^2 - k^2 + i \epsilon)^2} J_0(k) J_0 \left(k \frac{r}{R} \right) \nonumber \\
- &\lim_{\epsilon \rightarrow 0^+} \frac{r}{R} \int_0^\infty d k \frac{1}{m_a^2 R^2 - k^2 + i \epsilon} J_0(k) J_1 \left(k \frac{r}{R} \right) \,,
\end{align}
and evaluate the integrals by making use of \cite[p.~429~(1)]{Bessel} and \cite[p.~462~(44)]{Bessel2}
\begin{align}
\lim_{\epsilon \rightarrow 0^+} \int_0^\infty d k \, &\frac{2k}{(m_a^2 R^2 - k^2 + i \epsilon)^2} J_0(k) J_0 \left(k \frac{r}{R} \right) \nonumber \\
&= - \frac{i \pi}{2} \frac{H_1^+(m_a R) J_0(m_a r) + \frac{r}{R} H_0^+(m_a R) J_1(m_a r)}{m_a R}
\end{align}
\begin{align}
\lim_{\epsilon \rightarrow 0^+} \frac{r}{R} \int_0^\infty d k \, &\frac{1}{m_a^2 R^2 - k^2 + i \epsilon} J_0(k) J_1 \left(k \frac{r}{R} \right) \nonumber \\
&= \frac{\pi}{2} \frac{r}{R} \frac{Y_1(m_a R) J_0(m_a r) - i J_0(m_a R) J_1(m_a r)}{m_a R}\,,
\end{align}
to get
\begin{align}
I(m_a,r,R) = \frac{1}{m_a ^2 R^2} - \frac{i \pi}{2} \frac{H_1^+(m_a R) J_0(m_a r)}{m_a R} \, .
\label{I2}
\end{align}
\end{enumerate}
\Cref{eq:A} follows from~\cref{I1} and~\cref{I2}.

\bibliographystyle{JHEP_improved}
\bibliography{./refs}

\end{document}